\documentclass[%
reprint,
superscriptaddress,
amsmath,amssymb,
aps,prl,
]{revtex4-2}
\pdfoutput=1
\usepackage{graphicx}
\usepackage{dcolumn}
\usepackage{bm}

\usepackage[utf8]{inputenc}
\usepackage[T1]{fontenc}
\usepackage{mathptmx}
\usepackage{xcolor}
\usepackage{textcomp}
\usepackage{gensymb}
\usepackage{siunitx} 

\newcommand{\um}{\,\si{\micro\meter}}

\newcommand{\uL}{\,\si{\micro\liter}}
\newcommand{\mM}{\,\si{\milli\textsc{m}}}

\begin{document}
	
	\preprint{APS/123-QED}
	
	\title{Tunable active rotational diffusion in swimming droplets}
	
	\author{Adrien Izzet}
	\affiliation{Center for Soft Matter Research, Physics Department, New York University, New York, NY 10003, USA}
	\author{Pepijn Moerman}%
	\affiliation{%
		Debye Institute for Nanomaterials, Department of Chemistry, Utrecht University, Netherlands}
		\author{Jan Groenewold}%
	\affiliation{%
		Debye Institute for Nanomaterials, Department of Chemistry, Utrecht University, Netherlands}
	\affiliation{%
	Guangdong Provincial Key Laboratory of Optical Information Materials and Technology and Institute of Electronic Paper Displays, South China Academy of Advanced Optoelectronics, South China Normal University, Guangzhou 510006, P. R. China.}
	\author{Jér\^ome Bibette}%
	\affiliation{Laboratoire Collo\"ides et Matériaux Divisés, CNRS UMR 8231, Chemistry Biology \& Innovation, ESPCI Paris, PSL Research University, 10 rue Vauquelin, 75005 Paris, France}
	\author{Jasna Bruji\'c}
	\email{email: \texttt{jb2929@nyu.edu}}
	\homepage{https://wp.nyu.edu/brujiclab/}
	\affiliation{Center for Soft Matter Research, Physics Department, New York University, New York, NY 10003, USA}
	
	
	\date{\today}
	
	\begin{abstract}

Here we characterize the motility of athermal swimming droplets within the framework of active rotational diffusion. Just like active colloids, their trajectories can be modeled with a constant velocity $V$ and a slow angular diffusion, but the random changes in direction are not thermally driven. Instead, $V$ is determined by the interfacial tension gradient along the droplet surface, while local micellar fluctuations lead to droplet reorientation with a persistence time $\tau$. We show that the origin of locomotion is the difference in the critical micellar concentration $\Delta$CMC in the front and the back of the droplet. Tuning this parameter by salt controls $V$ from $3-15$ diameters $d/s$. Surfactant concentration has little effect on speed, but leads to a dramatic decrease in $\tau$ over four orders of magnitude. The corresponding range of the effective diffusion constant $D_e$ extends beyond the realm of synthetic or living swimmers, in which $V$ is limited by fuel consumption and $\tau$ is set by temperature or biological activity, respectively. Our tunable swimmers are ideal candidates for the study of the departure from equilibrium to high levels of activity, on both the single particle level and their collective behavior, including the motility-induced phase separation (MIPS). 
      	\end{abstract}

	\maketitle

Most active particles, such as Janus colloids~\cite{Howse_PRL2007}, are anisotropic and their orientation determines the swimming direction. They consume fuel preferentially on one side of the particle, which results in chemical gradients that drive and direct self-propulsion. On short time scales, these swimmers move in a straight line, but they are small enough that thermal diffusion randomizes the swimming direction on long time scales. The transition between these two regimes is characterized by the persistence time, and they are modeled as Active Brownian Particles (ABPs)~\cite{CatesTailleur_EPJ2015,Cates_EPL2013}. 
Unlike Janus particles, active droplets are chemically isotropic, yet they spontaneously break symmetry as they dissolve to create swollen micelles~\cite{Weber_RepProgPhys2018}. At a sufficiently large Peclet number, the isotropic state becomes unstable to small perturbations, resulting in self-sustained motion in a given direction~\cite{Michelin_PoF2013}. Autochemotaxis of the droplet is explained by an interfacial gradient along the surface, which induces a Marangoni stress and drives the flow of surfactants. This general mechanism is true for all droplet swimmers: direct emulsions made of oil~\cite{Moerman_PRE2017} or nematic liquid crystals~\cite{Thutupalli_Langmuir2012, Maass_PRL2016, Ueno_Langmuir2017, Suga_PRE2018}, as well as inverse emulsions~\cite{ThutupalliHerminghaus_NJP2011,IzriDauchot_PRL2014}. 

In the case of inverse emulsions, adding salt reduces the dissolution rate of the water into the oil and slows down the droplets, while increasing surfactant concentration speeds them up~\cite{ThutupalliHerminghaus_NJP2011, IzriDauchot_PRL2014}. It is thought that droplet motion is sufficient to cause the asymmetry needed to sustain a surfactant gradient, as the droplet encounters empty micelles in the front and leaves filled ones at the back~\cite{Michelin_PoF2013,Maass_PNAS2017}.

\begin{figure}[ht!]
	\includegraphics[scale=0.25]{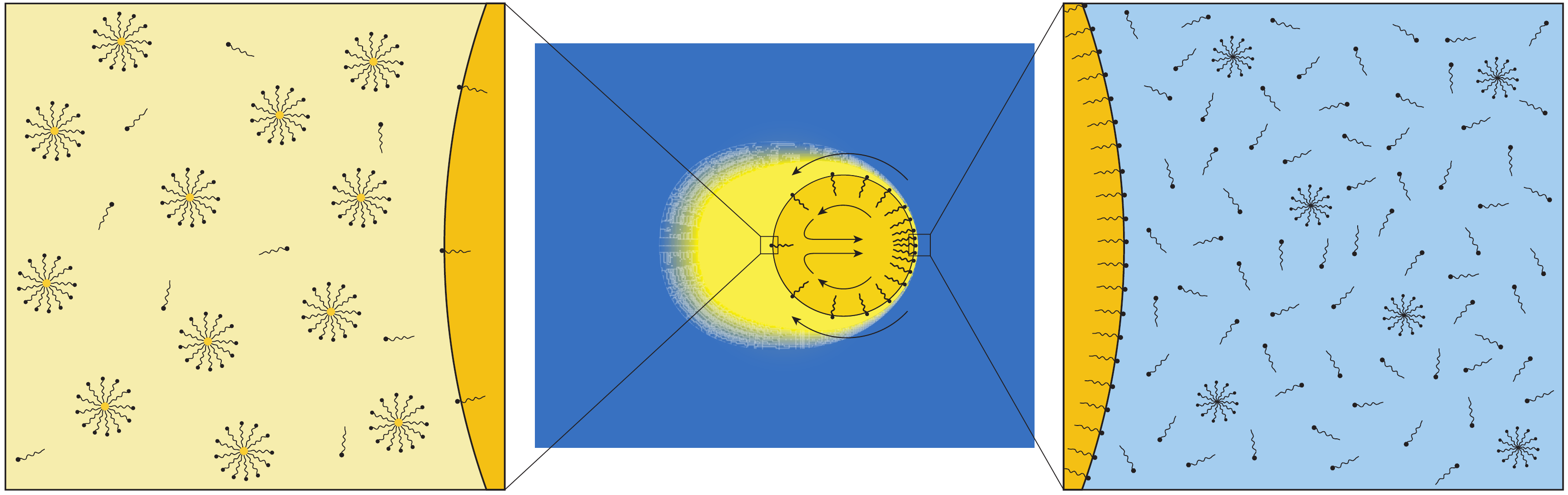}
	\caption{\label{fig:mechanismDeltaCMC} Schematic of the mechanism at the origin of self-sustained droplet motion. Spontaneous symmetry breaking arises due to a higher concentration of monomers in equilibrium with the aqueous solution (on the right) compared to those facing DEP-saturated water (on the left). This effect arises because of the difference in the CMC in the two environments, and it causes a Marangoni flow, shown by the arrows, which is enhanced by droplet motion.}
\end{figure}

In this letter, we present an alternative origin for the driving force of Marangoni flows in direct emulsions. The difference between the CMC of the surfactant in pure water and in the presence of oil, $\Delta$CMC, creates a concentration gradient of monomers from the front to the back of a droplet, as it swims and dissolves, as shown in the schematic in Fig.~\ref{fig:mechanismDeltaCMC}. The resulting Marangoni flow gives rise to and sustains speeds up to $15d/s$, an order of magnitude faster than any known swimming droplets. We demonstrate that adding salt decreases $\Delta$CMC proportionally to the swimming speed, validating the mechanism. Conversely, we show that adding surfactants above the CMC has little effect on speed, since $\Delta$CMC is independent of surfactant concentration.

\begin{figure}[t!]
	\includegraphics[scale=0.2]{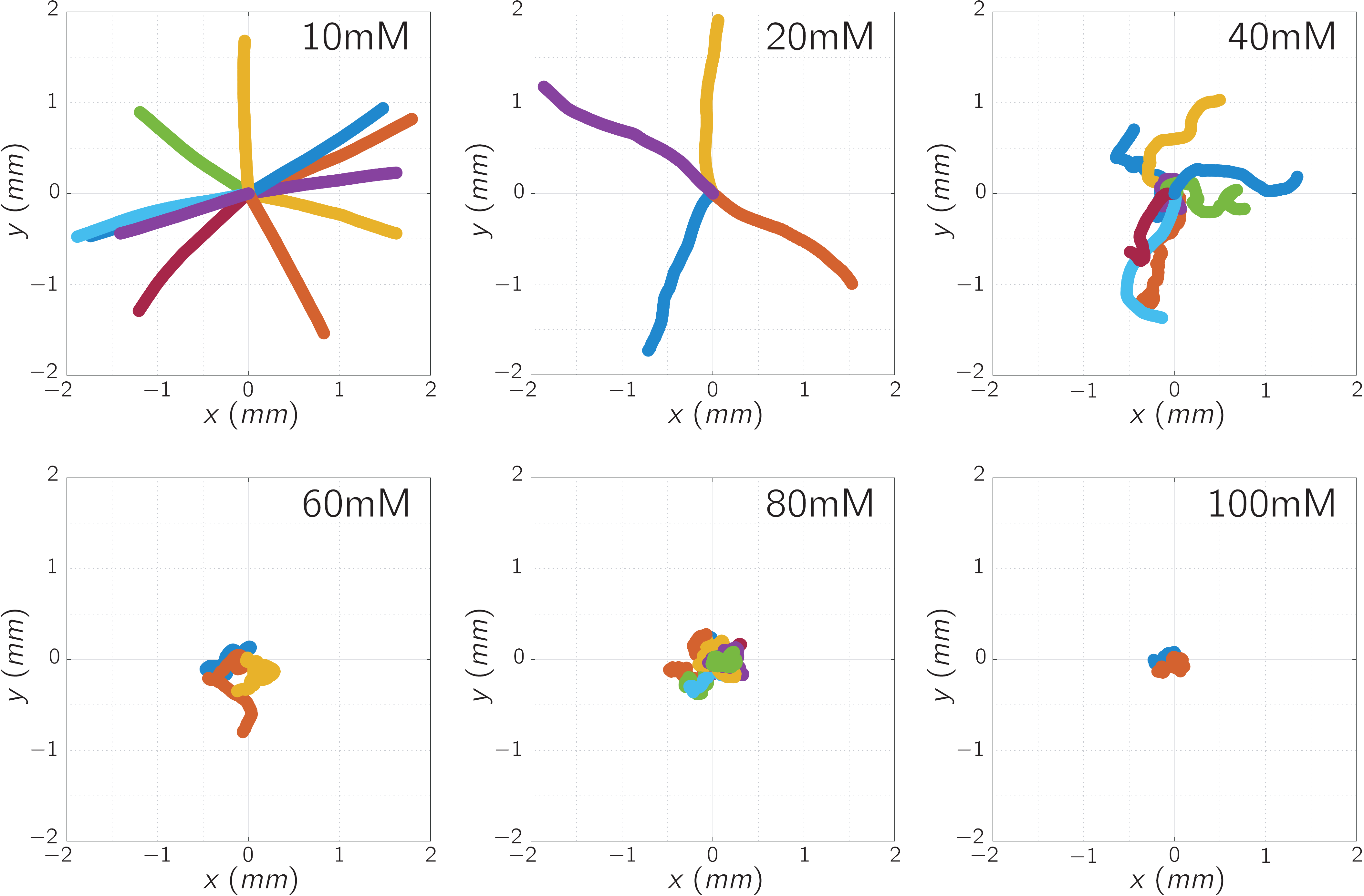}
	\caption{\label{fig:trajectories} Individual trajectories of $30$\um~diameter droplets, lasting four seconds, swimming in aqueous solutions with increasing bulk SDS concentrations. The motion transitions from ballistic to diffusive, with more frequent turns leading to a progressively lower effective diffusion constant, which is still orders of magnitude higher than that of thermal rotational diffusion.}
\end{figure}

The direction of droplet motion is fixed by the orientation of the surfactant gradient, which varies in time due to local micellar fluctuations, as seen in the trajectories in Fig.~\ref{fig:trajectories}. Because swimming droplets are athermal, ranging in size from $10-300\mu$m, their reorientation cannot be a thermally driven process and their motility is more complex than that of active colloids.

Depending on their size and fuel concentration, previous studies have empirically characterized trajectories of ordinary emulsions as ballistic (straight) or diffusive (random)~\cite{IzriDauchot_PRL2014, Suga_PRE2018}, and those of liquid crystal droplets as helical~\cite{Maass_PRL2016} and even oscillatory~\cite{Herminghaus_SoftMatt2014, Suga_PRE2018}. Despite all these advances, a quantitative analysis of the motility profiles as a function of experimental control parameters, as well as their theoretical interpretation, is still lacking in the literature. 

Here we show that single-particle dynamics in direct emulsions can be described by active rotational diffusion, in which velocity $V$ and persistence time $\tau$ are independently tuned by the salt and surfactant concentration, respectively. Since the microscopic mechanism that underlies turning is active rather than thermal, we are able to tune $\tau$ over four orders of magnitude. As a result, the effective diffusion constant is millions of times higher than that of passive rotational diffusion. Swimming droplets therefore belong to the class of ABPs, but with a much wider effective temperature range than active colloids or even bacteria~\cite{Buguin_PlosOne2012,Patteson_SciRep2015}. This system opens the possibility to study collective motion, MIPS, and giant fluctuations, which are predicted to occur at very high levels of  activity~\cite{Krauth_NatureCom2018}. While the ABP model holds for small droplets $< 50$\um, we note that more exotic effects, including self-avoidance~\cite{Maass_PNAS2017} and oscillations~\cite{Suga_PRE2018}, are observed for larger emulsions, and are beyond the scope of this study.     

Droplets of diethyl phthalate (DEP) are produced in a custom-built glass microfluidic setup with a coaxial co-flow geometry ~\cite{Cramer_ChemEngScience2004,  Choi_AdvMaterials2013, Haase_AngewandteChem2014}. The continuous phase is initially a $10$mM sodium dodecyl sulphate (SDS) solution saturated with DEP to prevent autophoretic motion and stabilize the droplets. To initiate swimming, $0.5$\uL~of the emulsion is injected using a micropipette into a circular glass sample cell ($1$cm wide and $1$mm deep), filled with SDS solution at the desired concentration. Droplet trajectories are observed using a Nikon Eclipse T\emph{i} microscope equipped with a $4\times$ Plan Apochromat $\lambda$ Nikon objective ($NA=0.20$). The movies are recorded using a Zyla sCMOS Andor camera operating at $40$ frames per second. To compare droplets of the same size, we crop the trajectories in a time window corresponding to a droplet diameter of $30 \pm 5$\um{}.

Our system therefore consists of droplets of DEP suspended in an aqueous solution of SDS. This oil is slightly soluble in water (1mM), which lowers the CMC of SDS from 8mM in pure water to 5mM in the vicinity of the droplets. This decrease corresponds to a decrease in $\Delta G_\text{mic} = 2.3 \, k_B T \, \log{\left( \Delta \text{CMC} \right)}=1.1k_B T$ for the micellization energy of one monomer of SDS~\cite{Israelachvili2011,RosenKunjappu2012}. The resulting DEP swollen micelles are thermodynamically more stable than the empty ones. 
Therefore, an infinitesimal perturbation in the velocity of the droplet leads to an anisotropy in surfactant concentration, on the order of $\Delta$CMC, on the droplet interface: the part facing DEP-saturated water sheds more surface active monomers into solution than the part facing the empty micelles in the bulk, as shown in Fig.~\ref{fig:mechanismDeltaCMC}. This asymmetry induces a Marangoni stress that forces the surface flow of surfactants. At high enough dissolution rates compared to molecular diffusion, the interfacial gradient undergoes a swimming instability and the droplets self-propel~\cite{Michelin_PoF2013}.

\begin{figure}[t!]
	\includegraphics[scale=0.45]{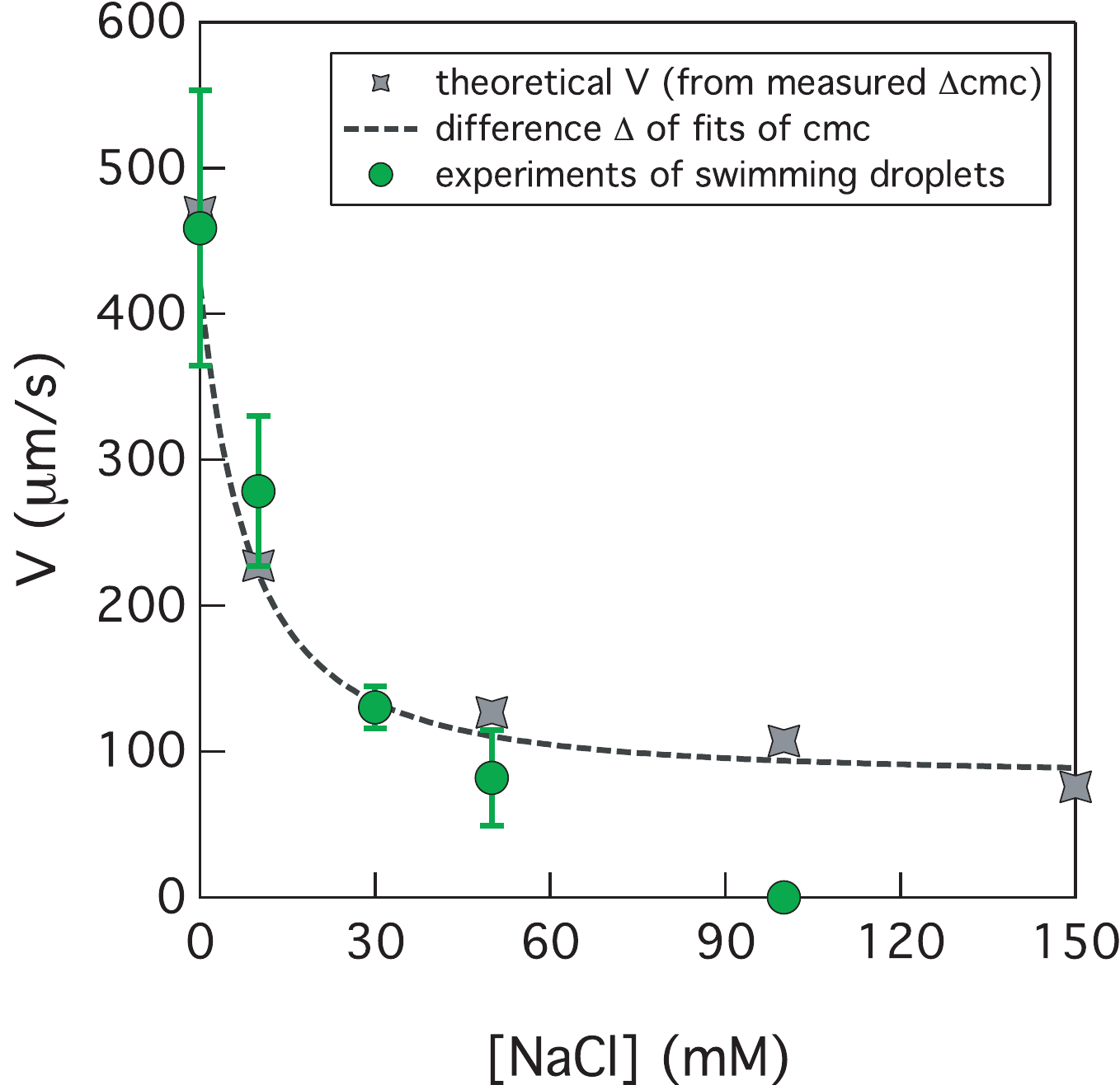}
	\caption{\label{fig:addingNaCl} Instantaneous speed $V$ at a constant $\text{[SDS]} = 10$\mM~sharply decreases with $\text{[NaCl]}$. This trend is theoretically captured by Eq.~\ref{eq:mobility} with $\Delta \text{CMC}$ values independently measured by pendant drop experiments. Above $\text{[NaCl]}=70$\mM, corresponding to $\Delta \text{CMC}=0.5\mM$, droplets cease to swim. Dashed line is computed from the difference between empirical fits of CMC vs [SDS]~\cite{Morton2008}).}
\end{figure}

More quantitatively, the droplet velocity is given by $\vec{V}=M \, \nabla_{||} C$, where $M$ is the phoretic mobility and $\nabla_{||} C$ is the average concentration gradient along the surface of the droplet.  It has been shown that Marangoni effects are dominant compared to diffusiophoresis in swimming droplets~\cite{Anderson_AnnRevFluidMech1989}. The velocity is thus given by:
\begin{equation}
	\vec{V} = \frac{r}{2\eta_0 + 3\eta_i} \left( - \, \frac{\partial \gamma}{\partial c} \right) \nabla_{||} C,
\label{eq:mobility}
\end{equation}
where $r=15\um$ is the radius of the droplet, $\eta_i=13$mPa.s is the viscosity of the DEP oil, $\eta_o=0.89$mPa.s is the viscosity of the solute, $\partial \gamma/\partial c$ is the rate of change of surface tension with SDS concentration at the interface, and $\nabla_{||} C = \Delta\text{CMC}/{2r}$ is controlled by $\Delta$CMC. 

Using the pendant drop method, we measure the surface tension as a function of surfactant concentration, in the presence and absence of DEP, and find the difference in the CMC to be $\Delta\text{CMC}=$3mM~\footnote{See Supplementary Material.}\setcounter{footnote}{1}. For the experimentally determined speed $V\sim450$\um/s, shown in Fig.~\ref{fig:addingNaCl}, we then calculate $\partial \gamma/\partial c=0.01$N/m/M. Comparing this value to the tension gradient in the range of $\Delta\text{CMC}$ at the interface with pure water~\cite{Note1}, we find that it is two orders of magnitude smaller than the expected value. This discrepancy is consistent with the numerical prediction of Michelin~\cite{Michelin_PoF2013} that one can only achieve $1-10\%$ of the theoretical swimming speed in the Janus limit, depending on the Peclet number that relates the advective and diffusive surfactant transport. Alternatively, if the process is modeled as an adsorption at the interface, such that $-\nabla \gamma=k_B T\lambda \nabla C$~\cite{Anderson_AnnRevFluidMech1989}, the measured speed predicts an interaction length $\lambda=5$nm, which is in good agreement with the diameter of a SDS micelle of $3.7$nm~\cite{Hayashi_JPhysChem1980, Duplatre_JPhysChem1996}, as proposed by~\citeauthor{IzriDauchot_PRL2014}.       
                 
To test the hypothesis that the driving force for propulsion is $\Delta$CMC at the interface, we add sodium chloride (NaCl) to decrease the drive~\cite{Bibette_book2002} and measure the resulting instantaneous velocity at a fixed surfactant concentration of $10$mM. First, we estimate $\Delta\text{CMC}$ by fitting the surface tension as a function of SDS and salt concentration using the equation proposed in~\cite{Szyszkowski1908} to reveal a gradual decrease in the CMC in both pure water and DEP-saturated water~\cite{Note1}. Nevertheless, the difference in the CMC decreases because the CMC decays faster in pure water. 

Assuming all other terms in Eq.~\ref{eq:mobility} remain the same, the change in $\Delta$CMC accurately predicts the sharp decrease in the average instantaneous speed shown in Fig.~\ref{fig:addingNaCl}. This assumption is corroborated by the fact that the gradients $\partial \gamma/\partial c$ below the CMC are largely independent of salt~\cite{Note1}. If we consider the surface adsorption picture, results are also consistent since the micelle size, which sets the characteristic length scale $\lambda$, hardly changes as a function of salt~\cite{Duplatre_JPhysChem1996}. An important difference between experiment and theory is that droplets do not swim below $\Delta\text{CMC}\approx0.5$\mM, while the theoretical prediction approaches a constant speed. This result shows that the surfactant gradient achieved by motility alone is not sufficient for self-sustained droplet motion, contrary to previous belief. Direct control of the speed via $\Delta\text{CMC}$ not only presents a useful toolbox, but also lends support to the idea that micellar equilibration at the interface is at the origin of Marangoni flows.

\begin{figure}[t!]
		\includegraphics[scale=0.5]{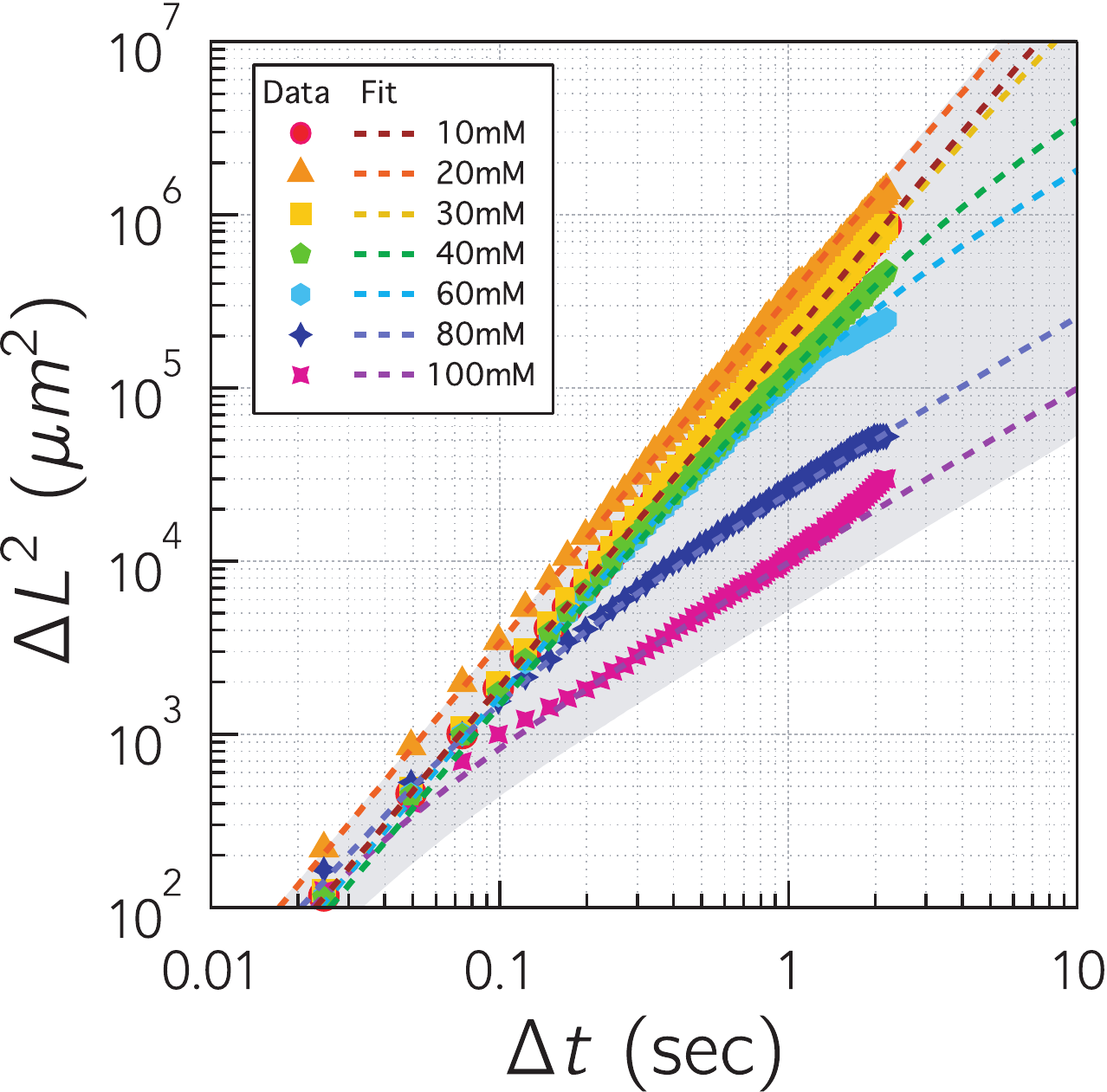}
		\caption{\label{fig:MSDfitRotDiff} Mean square displacement curves, averaged over tens of trajectories, show a decrease in the crossover time $\tau$ as a function of SDS concentration.  The model of rotational diffusion in Eq. ~\ref{eq:MSD} gives an excellent fit to the data and the grey area corresponds to the $95\%$ confidence interval.}
	\end{figure}

\begin{figure*}[t!]
	\includegraphics[scale=0.32]{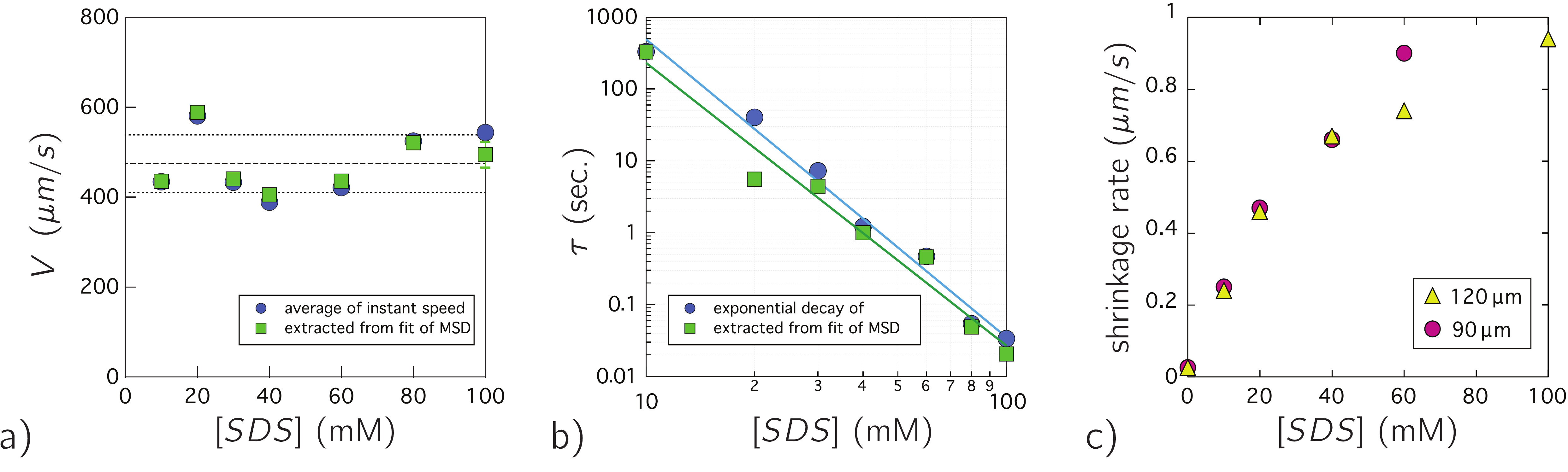}
	\caption{\label{fig:VandTau} Effect of surfactant concentration on motility. a) Instantaneous speed $V$, estimated directly from trajectories (circles) or the MSD fits of Eq.~\ref{eq:MSD} (squares), is largely independent of [SDS]. b) Persistence time $\tau$, from the fits and the values extracted from the velocity correlation function $C(\Delta t)$, sharply decays with [SDS] with a heuristic power law fit $\tau=6\times10^6\,$[SDS]$^{-4}$. c) The rate of droplet dissolution increases nonlinearly as a function of SDS concentration over the experimental range.}
\end{figure*}

Next, we investigate the effect of adding surfactants on the motility of individual droplets. Representative trajectories over four seconds are shown for various SDS concentrations in Fig.~\ref{fig:trajectories}. Slightly above the CMC, droplets swim with a constant speed in a given direction. This speed is only approximately constant since droplets decelerate at a rate of $10\mu$m/s$^2$, but this effect is negligible over the course of the trajectory. As the concentration increases, droplets change their direction more frequently, and therefore explore less space. To quantify the motility, we compute the mean square displacement (MSD) $\Delta L^2$ of the trajectories, shown on a log-log plot in Fig.~\ref{fig:MSDfitRotDiff}. We find that swimming droplets are ballistic at short times, where the slope is two, and they become diffusive at long times, where the slope is one. The crossover time $\tau$ between the two regimes decreases with increasing SDS concentration.

These results can be fit by the persistent random walk model~\cite{Marchetti_PRL2012},
\begin{equation}
\Delta L^2 = 2 \tau^2 V^2 \left( \frac{\Delta t}{\tau} +  \exp{\left( -\frac{\Delta t}{\tau} \right)} -1  \right), 
\label{eq:MSD}
\end{equation}
where the fitting parameters $V$ and $\tau$ are shown in Figs.~\ref{fig:VandTau}a,b. We find that $V$ is largely independent of surfactant concentration, and in good agreement with direct measurements of the average instantaneous speed along the trajectories, as well as previous measurements on small droplets~\cite{Suga_PRE2018}. The speed is therefore also independent of the dissolution rate, which increases with SDS concentration, as shown in Fig.~\ref{fig:VandTau}c. This result suggests that the equilibration timescale of the monomers at the interface with the bulk is faster than micellar solubilization. Droplet locomotion is governed by $\Delta\text{CMC}$ and not the bulk micellar concentration, in support of the proposed mechanism for propulsion.

\begin{figure}[t!]
	\centering
	\includegraphics[width=1\linewidth]{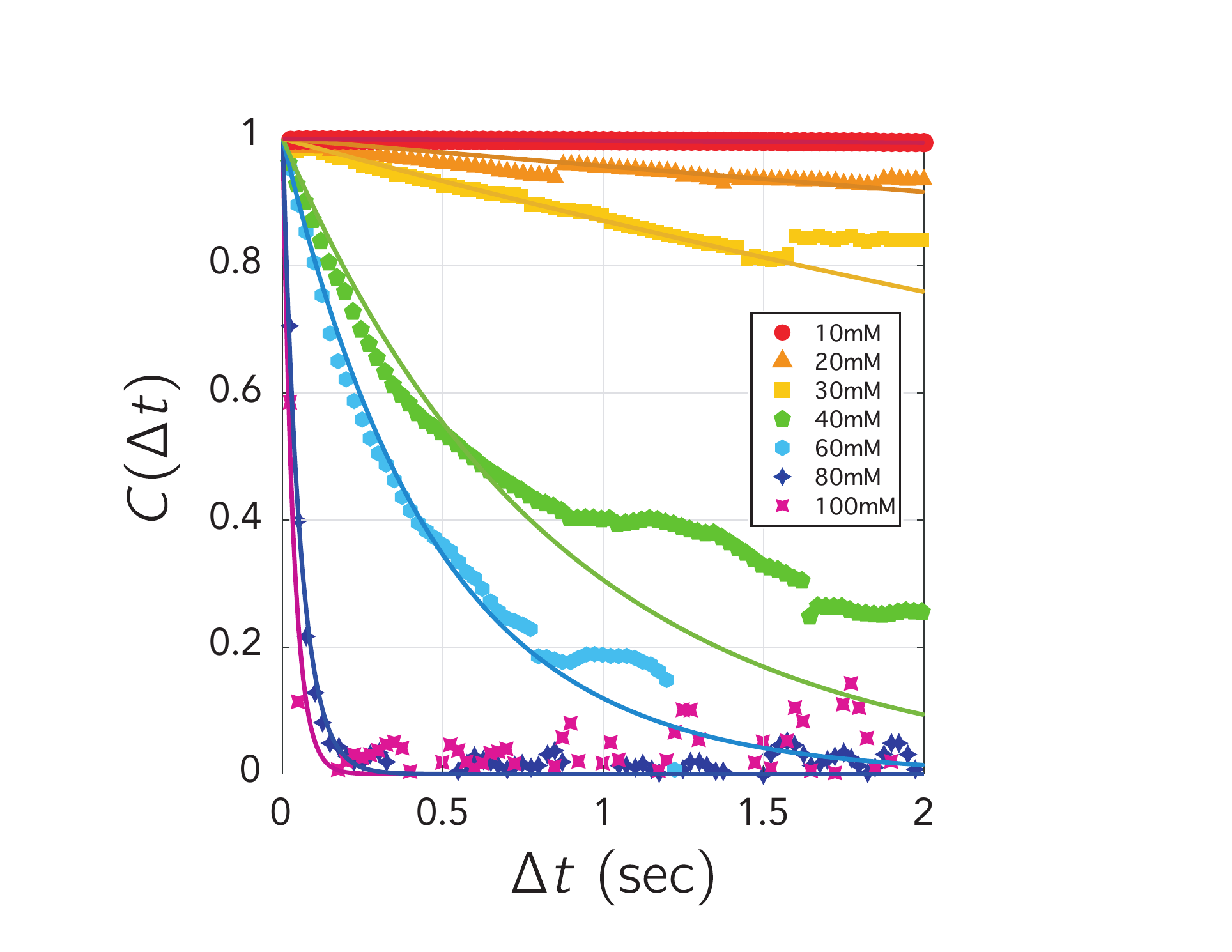}
	\caption{Normalized auto-correlation of the velocity for each concentration. Each dataset is computed from averaging the $C(\Delta t)$ of experimental trajectories at a given concentration. Solid lines are fit with an exponential decay following equation~\ref{eq:VCF_def}.}
	\label{fig:velocorrfitfinallegend}
\end{figure}

The surfactant concentration causes a decrease in $\tau$ that is dramatic: over one decade of SDS concentration, the fitting parameter $\tau$ in Eq.~\ref{eq:MSD} decreases over four orders of magnitude, giving rise to a power law exponent of $-4$ in the fit in Fig.~\ref{fig:VandTau}b. We compare the persistence time with the characteristic decorrelation time of the velocity-velocity correlation function $C(\Delta t)$ in Fig.~\ref{fig:velocorrfitfinallegend}, averaged over the whole trajectory~\cite{Burov_PNAS2013},
\begin{equation}
\label{eq:VCF_def}
C(\Delta t)=\bigg \langle \frac{\vec{v} (t)\cdot \vec{v}(t+ \Delta t)}{|\vec{v} (t)| |\vec{v}(t+\Delta t)|} \bigg \rangle_{t} .
\end{equation}
The data is fit reasonably well with an exponential decay, $C(\Delta t)=\exp(-\Delta t/\tau)$, to obtain the characteristic times over which the swimming direction randomizes, also shown in Fig.~\ref{fig:MSDfitRotDiff}b. The agreement between the two estimates of $\tau$ confirm the trend that droplets turn more frequently as more surfactant is added. This result differs from other swimmers, in which the persistence time is tuned by temperature, i.e. particle size~\cite{Palacci_Science2012}, or by biological processes in the case of living organisms~\cite{Patteson_SciRep2015}. This limits the range of reported $\tau$ to $0.1-1$s for colloids with thermally activated particle sizes. The measured active rotational diffusion constants $D_e=V^2\tau$ are on timescales that are at least one million times faster than thermal diffusion with $D\approx10^{-2}$\um$^2$/s.  

\begin{figure*}[ht!]
	\includegraphics[scale=0.5]{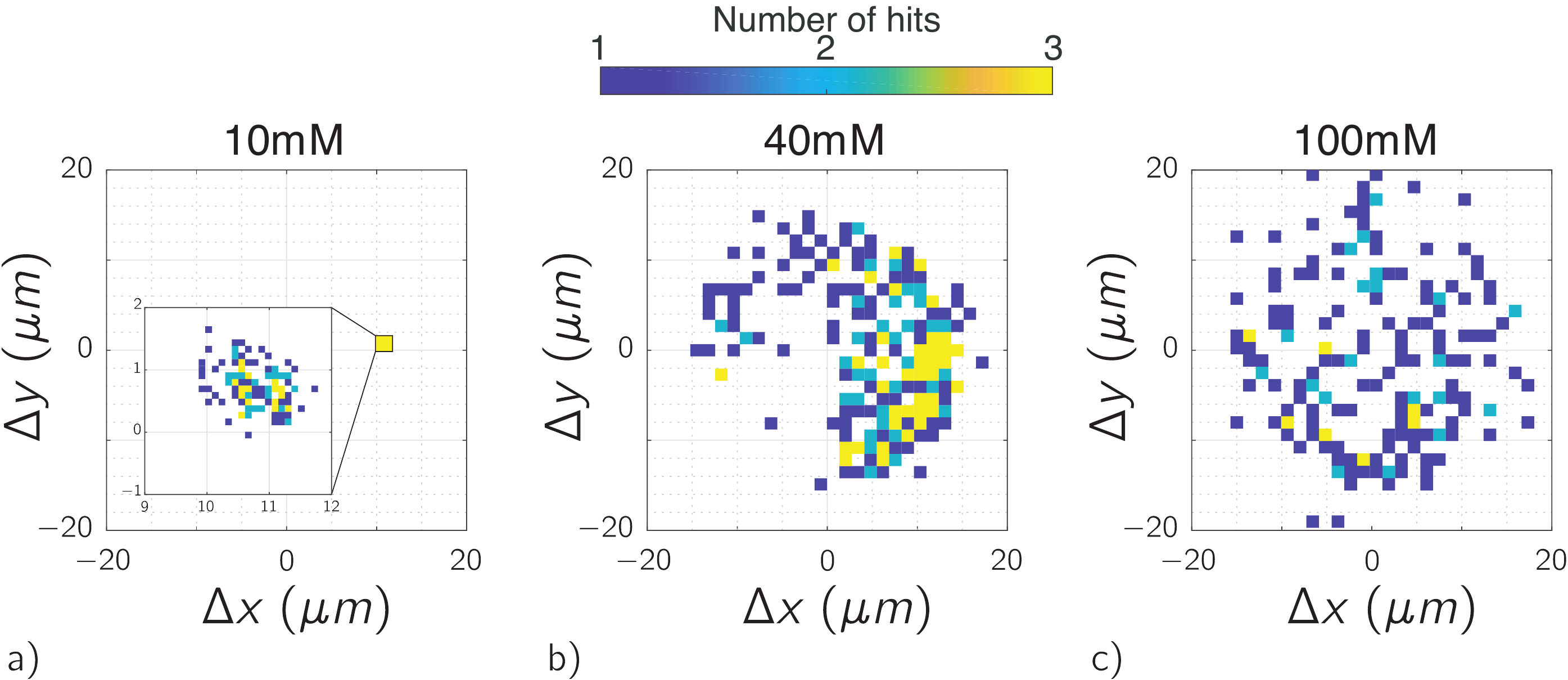}
	\caption{\label{fig:Displacement} Displacements of the droplet between consecutive time-steps (\emph{i.e.} $\Delta t =25$ms), for three surfactant concentrations. Colors represent the density at which this location in the displacement field is hit along the trajectory.}
\end{figure*}

The statistics of this active turning process are captured by the distributions of the instantaneous speed and reorientation angle as a function of SDS concentration in Fig.~\ref{fig:Displacement}. For a given $\delta t=$25ms, the displayed position at time $t+\delta t$ along the trajectory informs on both the speed and change in direction with respect to the previous timestep. Since the position at time $t$ lies at the origin, positions on the positive x-axis indicate straight motion, while particles that turn result in an off-axis position. As the SDS concentration increases, the characteristic time $\tau$ decreases to approach $\delta t$. When $\tau\gg\delta t$, we observe ballistic trajectories focused on a single point in panel (a), where the noise is on the order of a pixel size. Intermediate concentrations give rise to noise in both speed and reorientation angle, but the droplets rarely turn back on themselves (panel (b)). Eventually, at $\tau\approx\delta t$ droplets sample the space with equal probability,similar to Brownian diffusion, as indicated by the color map in panel (c). 

To explain this active rotational diffusion, we estimate that the increase in local micellar fluctuations surrounding the droplets creates gradients that are insufficient to reorient the droplets according to Eq.\ref{eq:mobility}. Instead of a microscopic mechanism for turning, it has been shown theoretically and numerically that a shift in motility from a Marangoni flow to diffusiophoresis~\cite{Schmitt_PoF2016, Morozov_JChemPhys2019,ChamolliLauga_EPJE2019} gives rise to the spontaneous breaking of symmetry of the inner and outer flow fields surrounding the droplet, eventually leading to chaos in the hydrodynamics of the droplet. Further studies on the spatial heterogeneities in the advective flows are needed to rationalize the turning frequency characterized by the rapid power law decay in the persistence time in Fig.~\ref{fig:MSDfitRotDiff}b. 

In this paper, we have presented evidence that the difference in the CMC at the front and back of the droplet is at the origin of droplet motion, which is why the average instantaneous velocity is independent of surfactant concentration above the CMC, but sensitive to changes in salt. Conversely, the frequency with which droplets change direction drastically increases with surfactant concentration, allowing us to independently tune $V$ and $\tau$ in the experiments. This allows one to achieve effective diffusion constants that are millions of times higher than those induced by passive rotational diffusion of active particles. These swimmers broaden the range of active particle sizes to the athermal realm. Such a system promises to be useful in tuning the departure from equilibrium~\cite{Gonnella_CRPhysique2015,Fodor_PRL2016} and understanding the resulting dynamics~\cite{CatesTailleur_EPJ2015,Stark_AccChemResearch2018,Thutupalli_PNAS2018}. The self-assembly of droplets into equilibrium colloidomers~\cite{McMullen2018} can now be extended to the assembly of non-equilibrium architectures, akin to living systems~\cite{Zeravcic2017}. 

The authors gratefully acknowledge insights by Eric Vanden Eijnden, Katherine Newhall,  Alexander Grosberg, Sascha Hilgenfeldt, Willem Kegel, and Alfons Van Blaaderen,  as well as useful conversations with Micha Kornreich and Rodrigo Guerra.
This work was supported by the Materials Research Science and Engineering Center (MRSEC) program of the National Science Foundation under Grants No. NSF DMR-1420073, No. NSF PHY17-48958, and No. NSF DMR-1710163. 
\bibliographystyle{apsrev4-2}
\bibliography{Swimmer_bibliography}
\end{document}